\documentclass [aps,pra,amssymb,amsmath,showpacs,twocolumn]{revtex4-1}
\usepackage[latin9]{inputenc}
\usepackage{amsmath}
\usepackage{amssymb}
\usepackage{graphicx}
\usepackage{verbatim}
\usepackage{bm}
\usepackage{color}

\usepackage{graphicx,epsfig,amsfonts,amssymb}
\usepackage{bm}
\usepackage{times}
\usepackage{lipsum}
\usepackage{verbatim}

\newcommand{\la}{\langle}
\newcommand{\ra}{\rangle}

\newcommand{\be}{\begin{eqnarray}}
\newcommand{\ee}{\end{eqnarray}}

\newcommand{\bs}{\begin{equation}\begin{split}}
\newcommand{\es}{\end{split}\end{equation}}

\date{\today}

\begin{document}
\title{Computing with a single qubit faster than the computation quantum speed limit}
\author{Nikolai A. Sinitsyn} 
\affiliation{Theoretical Division, Los Alamos National Laboratory, Los Alamos, NM 87545,  USA}

\begin{abstract}
 The possibility to save and process information in fundamentally indistinguishable states is the quantum mechanical resource that is not encountered in  classical computing.
 I demonstrate that,  if energy constraints are imposed, this resource  can be used to accelerate information-processing without relying on entanglement or any other type of quantum correlations.  In fact,
  there are  computational problems that can be solved much faster,  in comparison to currently used classical schemes, by saving intermediate information in nonorthogonal states of just a single qubit.
There are also error correction strategies that protect  such computations.  
\end{abstract}

\maketitle

\begin{figure}
\includegraphics[width=5.0 cm]{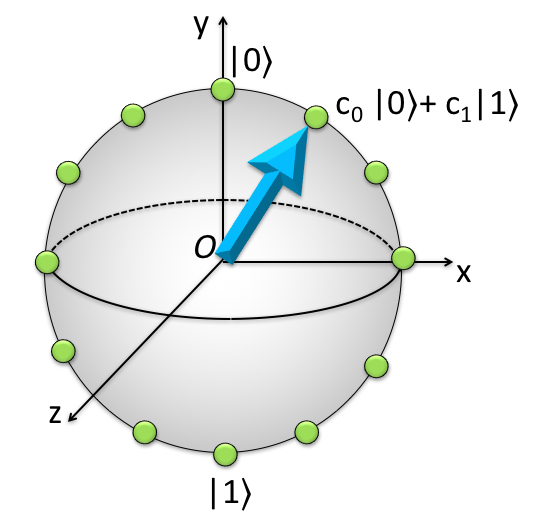}
\centering
\caption{Up to overall phases that do not influence measurement outcomes, states of a qubit correspond to points on the 2D sphere. This phase space can be discretized to create a register of states  (green circles) for computation. However, only opposite points on this sphere, such as the poles  marked by $|0\ra$ and $| 1\ra$, are distinguishable by measurements. }
\label{sphere}
\end{figure}

{\it Introduction.} The quantum phase space of a qubit is a sphere (Fig.~\ref{sphere}). One can discretize this space into any number of states and
then apply field pulses to switch between the chosen states in an arbitrary order.  In this sense, a qubit  comprises the whole universe of choices for computation.  For example, a qubit can work as  finite automata \cite{q-auto} when different unitary gates act on this qubit depending on arriving digital words. 
However, different states of a qubit are generally not distinguishable by measurements. So, if the final quantum state encodes the result of computation, we cannot generally extract this information because we cannot  distinguish this state  by a measurement from other non-orthogonal possibilities reliably. 

For such reasons, qubits are believed to provide computational advantage over classical memory  only when they are used to create purely quantum correlations, i.e., entanglement or quantum discord \cite{qdiscord}. While very powerful algorithms have been designed based on such correlations,  the degree of control over the state of many qubits that is needed to implement commercially competitive quantum computing is far from the level of the modern technology.

In this note, I will argue that the ability to use non-orthogonal states for computation should be considered as the completely independent resource  that is provided by quantum mechanics. With a specific example, I will show that there are computational problems for which the access to just one 
high quality qubit may  provide speed of computation that, fundamentally, cannot be reached by a classical computer under the specified restrictions on  raw resources such as memory coupling strength to control fields.

 The idea of this article is based on the well known observation that time-energy uncertainty relation in quantum mechanics imposes limits on computation speed at fixed power supply for classical schemes of computer operation \cite{margulis,lloyd}. Such claims are generally justified by the fact that digital computers save information in the form of clearly distinguishable states, such as $0$ and $1$ that encode one bit of information. Quantum mechanically, distinguishable states must be represented by orthogonal  vectors that produce definitely different measurement outcomes. However, the switching time between two orthogonal quantum states is restricted  from below by a fundamental computation speed  limit $T=h/(4\Delta E)$, where $\Delta E$ is  characteristic energy of  the control field  coupling to the memory that is needed to update one  bit of information \cite{vaidman}. 
 So, restrictions on strength of control fields automatically restrict the speed of classical computation that saves information in physically distinguishable states. 
While the existence of this {\it computation quantum speed limit} is a mathematically proved fact, I will show an explicit elementary example that demonstrates possibility of solving a computational problem  faster than the lowest time bound that is imposed by this limit on classical computation hardware. Access to the quantum memory makes this possible because it allows information-processing using nonorthogonal quantum states. So, there is no more direct linear relation between 
the minimal time and the number of elementary logic operations required to implement an algorithm at given energy constraints.

{\it The remainder of the Hamming weight.} 
If there is  only a single available qubit for computation, there is no possibility to discuss quantum correlations. However,  one can still perform switches between state vectors of this qubit that are arbitrarily close to each other in the phase space. At fixed strength of control fields this  can be done much faster than switching between orthogonal states. The question is only whether we can effectively read the result of such manipulations in order  to solve a legitimate computational problem.

Imagine that our computer receives a long string of zero and unit numbers: 
\be
1,1,0,0,1,0,1,\ldots, 0, 1,1,
\label{string1}
\ee
with the total number of $N\gg1$ characters. 
 The number $N_1$ of unit characters in such a string is called the Hamming weight. We assume that it  is unknown, and computer has the task to answer the question: ``which one of the integer numbers, $0$ and $n$, is {\it not} the remainder of division of $N_1$ by $2n$?".
 This is a legitimate computational problem.  For example, let $2n=4$ and  $N_1=1729 = 432\cdot 4+1$. Since the remainder is $1$ then neither $0$ nor $2$  is the remainder. Hence, if machine returns either $0$ or $2$ it gives a correct answer in this case. Another example: $N_1=8=2\cdot 4+0$. The remainder is $0$ so the only correct answer that machine must return should be $2$.

Let me consider that $N> n > 1$ and estimate the minimal time and hardware resources that are needed to solve this problem classically. Suppose that characters of the string are processed with constant time intervals $\tau$ per character, while units are separated by unknown chains of zeros. 
Each time a unit number arrives, we must update our records. Since only remainder of division by $2n$ matters, we need only $\log_2(2n)$ classical memory bits that should be updated to keep information about the remainder of the division of the already arrived number of units by $2n$.  
Since classically distinguishable states must be quantum mechanically orthogonal, each flip of a memory bit should be induced by a pulse with field-memory coupling energy  $\Delta E\ge h/(4\tau)$. This is the energy cost of counting each unit character classically. Conversely, if our computer has restrictions on the strength of control fields that it can create, the time of processing one character has to be restricted as $\tau \ge h/(4\Delta E)$, so the total time of computation is fundamentally restricted as 
\be
T\ge Nh/(4\Delta E).
\label{time}
\ee

Quantum mechanically, the same computational problem can be solved with only a single qubit. Indeed, let us assume that this qubit is  a spin-1/2, which points up along $y$-axis initially, i.e., it is in the state $|0\ra$ in Fig.~\ref{sphere}. Each time a unit character arrives, it triggers a magnetic field pulse, along the $z$-axis,  that rotates the spin counterclockwise  by an angle $\pi/n$ in the $xOy$ plane. 
Remainders $\{ 0,1,2, \ldots 2n-1 \}$ are then encoded in spin rotation angles, respectively, $\{ 0, \pi/n, 2\pi/n, \ldots (2n-1)\pi/n \}$. Note that we identify rotation angles that are different by multiples of $2\pi$ because they represent the same spin state vector. 
 
 After the full string of characters passes through such a computation, we perform measurement by a projection operator on the state with zero rotation angle: 
 \be
 \nonumber \hat{X}=|0\ra \la 0 |.
 \label{proj1}
 \ee
Although the spin states that encode possible remainders are generally not  orthogonal, particular states that represent remainders of interest, $0$ and $n$, are represented by quantum mechanically orthogonal states with spin rotation angles, respectively, $0$ and $\pi$.  Suppose the outcome of measurement is $X=1$. This outcome is possible  for all possible spin rotation angles except $\pi$. 
 So,  receiving $X=1$ we will conclude that the number $n$ is definitely not the remainder of division of $N_1$ by $2n$. In the alternative case when the measurement outcome is $X=0$, we will conclude that $0$ is definitely not the remainder of division of $N_1$ by $2n$. So, our task will be fulfilled.

Let me  now examine energetics of this computation. At each elementary step spin rotates by  an angle $\pi/n$, which is $n$ times smaller than what is needed to switch between orthogonal spin states. Repeating standard arguments, e.g. from  Ref.~\cite{tamm,vaidman},  I find that such an elementary operation  requires coupling energy that is limited by $\Delta E=h/(4n\tau)$, i.e., $n$ times less than what is needed for switching between  orthogonal states. For spin-1/2, this limit is reachable because it is achieved by a square pulse of a constant magnetic field along the $z$-axis. 
Consequently, by using the quantum memory we reduce the coupling between the field and the memory in our processor by the factor $n$ or, equivalently, speedup calculations $n$ times at fixed strength of this coupling. 

The most time and energy consuming step is the final measurement. It is done only once and therefore does not influence scaling of the performance of the algorithm with $N$. Moreover, any other computation scheme would require at least one such a measurement to obtain the result. So, this step does not reduce the performance 
in comparison to classical schemes.

Finally, let me compare the qubit  to  a classical rotator in the same computation scheme, i.e., assuming the same discretization of the spherical phase space and the same switching protocols. A classical  rotator is physically realized by a magnetic grain with a large effective spin $S \gg1$.  
Although our algorithm requires discretization of the spin phase space into $2n$ different rotation angles, we do not assume that we have to make records of transient states during computation. We can even assume that $n>S$, so motion in   the continuous classical spin phase space does not lead to infinite energy requirement. 

 Coupling of the classical spin to the magnetic field  is described by the Hamiltonian $\hat{H}={\bf B} \cdot {\bf S}$. Characteristic energy of this coupling is  
$\Delta E = |B|S$. The  spin rotation frequency, $|B|/\hbar$, is independent of $S$, while $\Delta E \propto S$. 
 The classical spin switches between rotation angles $2S$ times slower at the same characteristic coupling $\Delta E$ than a qubit with spin $1/2$. 
Thus, replacing a qubit with a classical spin, while keeping the same scheme of computation, slows the computation speed  at fixed $\Delta E$ down.



\begin{figure}
\includegraphics[width=5.0 cm]{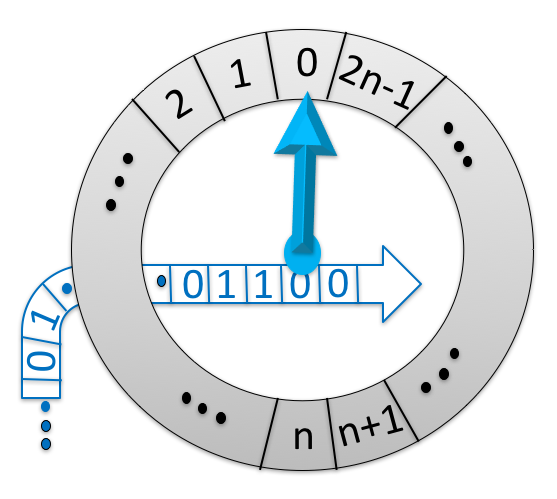}
\centering
\caption{A roulette with a pointer that chooses among $2n\gg1$ integer numbers changing from $0$ to $2n-1$. A large number with binary entries, $0$ and $1$, is randomly generated to rotate the pointer (marked by blue color).
Pointer arrow is initially directed to zero, and then it moves by an angle $2\pi/(2n)$ counterclockwise each time the input character ``1" is received.  The bet must be placed only either on  $0$ or on $n$. If at the end the arrow points to the number of the bet, the player looses everything. In all other cases, the player wins.}
\label{roulette}
\end{figure}

{\it Gambling example}. The considered computational problem may look quite artificial at first view.  However, the access  to one qubit computer  that implements the above algorithm   can actually give an advantage in realistic gambling games. Imagine the virtual online game with a roulette that has $2n$ discrete states, as in Fig.~\ref{roulette}. Stakes are received only on two states: $0$ and $n$. The pointer is initially set at zero and then rotates. If it ends on one of the numbers $0$ or $n$ then players who bet on that number {\it loose} everything. In all other cases they win a small amount, like in Hussar Roulette \cite{fatalist}.
 
Let me assume that to rotate a pointer a huge (many gigabytes) number of the type (\ref{string1}) is generated and then used bit-by-bit, so that when number $1$ arrives it triggers rotation of the pointer by the angle $\pi/n$. The number in Eq~(\ref{string1}) 
becomes available to public right before rotation of the pointer. Casino considers this safe because none of the classical computers can process this number before stakes are received. 
In this situation, the player that has access to a single qubit computer can cheat the game by simulating rotation of the pointer considerably faster, and thus has time to place the bet on the safe number.


{\it Error propagation and error correction}. The major two sources of errors that are expected for a spin qubit  are the finite fidelity of gates and the uncontrolled environmental magnetic field fluctuations. The former 
problem is more important because there are already many single and few qubit platforms with long coherence times \cite{deutsch,bechtold,dobrovitski,qubit1}. 
A non-perfect quantum gate can be characterized by the  typical difference $\phi_0$ between the rotation angle of a qubit and the desired angle $\pi/n$. Our computer can be tuned to make  systematic errors arbitrarily small but error variance will accumulate with the number of applied gates, so after processing the string the variance of angle deviation from the desired position is 
\be
{\rm var}(\phi)  = N_1 \phi_0^2.
 \label{dph1}
 \ee
 This variance also characterizes the probability of making the wrong decision in the critical situation that is  when the pointer of the roulette will end either at $0$ or at $n$.  
 For example, suppose the final state of the qubit  must  be $|0\ra$ but spin makes additional rotation by a small angle $\phi$. The state of the qubit is then $\cos(\phi)|0\ra +\sin(\phi) |1\ra$, and the probability that  the measurement by operator $\hat{X}$ gives wrong, i.e. unit, outcome is 
 \be
 P= \sin^2(\phi) \approx  {\rm var}(\phi).
 \label{prob}
 \ee
So, the whole algorithm would generally work if at the end of computation we achieve the condition 
\be
\phi_0\sqrt{N} \ll 1.
\label{conder}
\ee
The precision of the rotation angle should grow then as  $\sim \sqrt{N}$ with increasing the length of the string (\ref{string1}). 

Let me now assume that condition (\ref{conder}) is achieved at the hardware level but the probability (\ref{prob}) of making wrong bet is already not acceptable. 
Classical approaches to error correction are sufficient to resolve this problem.
For example, instead of one qubit, we can prepare three identical ones with identical initial conditions. One should then let all qubits go through the same sequence of field pulses, so they end up in similar states, with random uncorrelated deviations from the rotation angle $\phi = N_1\pi/n$. Only the critical situations when $\phi=k\pi$ with integer $k$ matter. Otherwise the player wins no matter what is the bet. So, let me assume that $k$ is even. The states of qubits are then given by 
\be
|u_i\ra = \cos(\delta \phi_i) |0\ra + \sin(\delta \phi_i) |1\ra, \quad i=1,2,3,
\label{qub}
\ee 
where $\delta \phi_i \ll 1$. Before measurements, the total state is 
\be
|\psi \ra = |u_1\ra \otimes |u_2\ra \otimes |u_3\ra.
\label{direct}
\ee
One can then apply measurement operator $\hat{X}$ to all the qubits at the end of computation. 
Simple majority voting, i.e., interpreting outcomes with one unit and two zeros as the final zero result would reduce the probability of wrong bet to the value $O([{\rm var}(\phi)]^2)$. So, the final  probability of an error in the critical situations can be reduced by orders of magnitude at expense of slowing down calculation by factor 3. This slowing is not essential in the case $N>2n \gg 1$.

It is also conceivable to perform  quantum error correction by  entangling the original qubit with several others and then using the larger phase space. Such error correction strategies can employ  ideas from quantum metrology. For example, imagine that we have $n_q$ high quality qubits and suppose that we can apply unitary operator $\hat{U}$ that transforms trivial states  $ |0\ra^{\otimes n_q} $ and $ |1\ra^{\otimes n_q}$ into
GHZ states, respectively, $|-\ra$ and $|+ \ra$,  where
\be
|\pm \ra = \frac{1}{\sqrt{2}} \left( |0\ra^{\otimes n_q} \pm |1\ra^{\otimes n_q} \right).
\label{GHZ}
\ee
We can then encode discrete states $k=0,\ldots, 2n-1$ from Fig.~\ref{roulette} in generally nonorthogonal states 
\be
|\psi_k\ra= \frac{1}{\sqrt{2}} \left( |0\ra^{\otimes n_q} - e^{-i\pi k/n} |1\ra^{\otimes n_q} \right).
\label{ghz2}
\ee
Switching from $|\psi_k \ra$ to $|\psi_{k+1} \ra$ is done with the Hamiltonian $\hat{H} = \Delta E  |1\ra    \la 1|$, where $|1 \ra$ is the state ``up" of one of the qubits. 
This step requires coupling $\Delta E=h/(4n\tau)$ where $\tau$ is the time bound on the switching operation.  This coupling strength does not depend on the number $n_q$ of entangled qubits, and it is the same as the minimal coupling needed to rotate a single qubit by an angle $\pi/n$ during the same time interval. 
 Hence, we can implement our algorithm using states  (\ref{ghz2}) without any loss in speed of switching between nonorthogonal states in comparison to the single qubit case.
 
  In order to perform the error correction, at the end of computation, we apply the inverse unitary operator $\hat{U}^{-1}$. 
 Calculations with ideal qubits would then encode remainders $0$ and $n$ of the division by $2n$ in the final states $ |0\ra^{\otimes n_q} $ and $ |1\ra^{\otimes n_q}$, respectively. Small errors during computation would lead to finite probabilities to detect other states at the end with generally mixed 0 and 1 states of different qubits. Interpretation of such outcomes can be done then using the same majority voting as in the above discussed classical error correction strategy. The latter steps slow down calculations by the amount of time that does not scale with $N$. So, finally, the quantum error correction  may appear superior over the classical one because  switches between nonorthogonal GHZ-like states do not slow calculations down for $n_q>1$   qubits in comparison to $n_q=1$ in $N \gg 1$ limit.  We note, however, that this quantum error correction protects only against specific random qubit-flip errors while operations with entangled states (\ref{ghz2}) can enhance the probability of the error with 
 all-qubit flips, which cannot be treated with majority voting. We leave this problem for future studies.

{\it Estimation for experiment}. 
Qubits based on hyperfine levels of trapped ions, such as $^{43}$Ca, have  $ 50$ seconds of coherence time and  fidelity of single qubit gates better than 99.99\%  \cite{qubit1}, which corresponds to $\phi_0<1\times 10^{-2}$.  According to (\ref{conder}), it is possible to process a string with $N\sim 10^{4}$ binary characters by applying radio-frequency pulses leading to the probability of the final error $P=0.1$, which can be reduced to  $\sim 0.01$ by using the error correction with three qubits. Thus, efficient computations based on non-orthogonal quantum states can be demonstrated with modern quantum technologies. 

Certainly, it can be difficult to benchmark the speed of such computations versus classical schemes because, e.g., none of the classical computers works with fields coupled to nuclear spins. However, such experiments can prove that 

(i) certain computational problems can be solved by requesting less time per elementary logic operation than the  fundamental lower bound on switching time between orthogonal states  at given maximal coupling of control fields to  memory; 

(ii) a single qubit, like a classical rotator, can replace the work of $\log_2(2n) \gg 1$ classical bits; 

(iii) all this can be done without the need to create quantum correlations.

{\it Related prior work}.
Before the conclusion, I would like to  comment on publications  that have previously expressed similar looking views. 

It has been suggested in \cite{hardy}  that one can use advantages of the qubit's continuous phase space  in order to find   the area of a field pulse whose area is constrained to rotate the qubit by angles $\pi k$, with unknown binary integer $k $, e.g., $k\in \{1,2\}$. Different parities of $k$  lead to orthogonal, i.e. distinguishable final states of the qubit, and hence can be determined with a single qubit. On the other hand,  authors of \cite{hardy} claimed that to  determine this parity with a classical digital computer one would have to explore a digitized continuous  signal, which  needs formally infinite number of classical bits.  So, we arrive at a paradox that a classical digital computer needs infinite resources to do this. 

However, the paradox in \cite{hardy} has loopholes. A classical bit is actually capable of providing binary information about a continuous pulse. One memory bit can be coupled to the field so that the bit flips with the rate that depends on the field amplitude that it senses.  
The number of this bit flips by the end of interaction  does contain some binary information about the whole pulse. Similarly, it is not forbidden by classical physics to assume that a classical bit is designed to change its states depending on the area of the field pulse with which it interacts.

 Indeed, in classical mechanics, a classical bit is idealization. Memory bits are always realized  physically by structures with a continuous phase space, e.g., magnetic grains that can  behave as controlled rotators in external magnetic fields. The magnetization vector of a grain can point in any direction.  
Classical magnetic grains can equally well,  as qubits, sense continuous field pulses, and they actually do and must do this in practice when they have to ``decide" whether to flip between states $0$ and $1$ represented by local energy minima  in their phase space. 

So, analog sensing of field pulses is actually used even in commercial digital computers. We know from our  experience that this does not lead to requests of infinite energy resources even when devices operate according to laws of classical physics.  Hence, either  classical computers already use the effect discussed in \cite{hardy} at large scale 
routinely (see also \cite{marg} for similar discussion), or the emergence of classical realism somehow removes the infinite resource requests from classical devices when they do not have to make records of intermediate states during interaction with a continuous signal.   
Moreover, to be consistent with laws of physics, we can also recall that the true field cannot be continuous in quantum mechanics. It changes in discrete portions (e.g., photons) that can be processed by a bit sequentially. 
All these arguments  raise doubts in practical usefulness of the quantum resource  discussed in \cite{hardy}. 

Despite this critic of Ref.~\cite{hardy}, the algorithm in the present article is essentially a variation on the theme \cite{hardy}. The algorithm, however,
is free from poorly justified assumptions about the nature of classical bits and fields. The key difference from \cite{hardy} is my claim that unambiguous benefits appear not as the larger size of the classical memory that a qubit can replace but rather as the higher speed of computation that the quantum memory allows at finite coupling  to the control fields. Indeed, arguments that show advantages of the algorithm apply equally well to the example in Ref.~\cite{hardy} with the pulse area sensing. A single classical spin can, as the qubit, sense the area of the magnetic field pulse. This undermines the need of infinite classical memory for classical sensing. However, the smaller the spin size we use, the smaller the coupling energy $\Delta E$ that is needed to rotate the spin at the same speed. So the unambiguous quantum advantage of using the qubit here is merely the possibility to reach the maximal possible speed of sensing  at given fixed coupling  of the sensor to the field.

Next, I would like to mention publications that have previously concluded that there is no limit on speed of quantum computation. Ref.~\cite{no-limit} argued  that speed of entanglement may not be restricted by the standard speed limit because entanglement is created without orthogonal transformations.
In Ref.~\cite{recent}, similar conclusions were reached with a specific example that uses entanglement of several qubits. Both these articles justify their conclusion by the fact that the work of a quantum circuit that is made of a set of standard quantum gates can be simulated by evolution with a Hamiltonian that 
requests less energy resources. While this is certainly useful to know for optimization of quantum gates, quantum simulators discussed in \cite{no-limit,recent} are not complete algorithms,  so the comparison could be unfair.  For example, it is unclear whether a quantum simulator allows  energy effective error correction strategy. In fact, it is not surprising that the gate-based computation takes more resources than the process that this computation quantifies. It is also unclear from Refs.~\cite{no-limit,recent} whether purely quantum correlations, such as entanglement  and quantum discord, are required for energy efficiency. Purely quantum correlations are currently very hard to control even on the level of  few qubits, so energy efficiency of dealing with such correlations cannot become important in the near future.

Summarizing, in contrast to all these prior work, the computation example that I have shown is, simultaneously, 

(i) an algorithm that works with digitized information; 

(ii) it is not a  quantum simulator or a model of a sensor;

(iii) efficiency of this algorithm cannot be matched by replacing the qubit with a classical rotator while keeping the same partitioning of the continuous phase space; 

(iv) this algorithm does not rely on quantum correlations.   


{\it In conclusion}, conventional quantum computing is designed to achieve superior performance over classical computing by employing multistate quantum superposition and entanglement. However, I showed that when energy constraints and the time of computation, rather than the number of elementary logic operations needed to solve a computation problem, are the primary variables for optimization, then the possibility of fundamentally faster switches between nonorthogonal states  should be treated as another viable quantum resource for faster computations.  


While massive controlled quantum entanglement is beyond the reach of modern technologies, present work shows that 
alternative approaches, such as algorithms based on  quantum  finite automata \cite{q-auto}, may lead to  computations that are fundamentally faster than classical  ones because   information  can be saved and processed at intermediate stages of quantum automata algorithms by using fundamentally indistinguishable quantum states. Currently, there are many high-quality single qubit platforms that can demonstrate superior performance of such computations.

 


{\it Acknowledgements}. I thank R. Somma for useful discussion. This work
was carried out under the auspices of the National Nuclear
Security Administration of the U.S. Department of Energy at Los
Alamos National Laboratory under Contract No. DE-AC52-06NA25396. I also thank the support from the LDRD program at LANL.

\end{document}